\documentstyle[aps,prl,twocolumn]{revtex}
\input epsf

\begin{document}
\draft
\title{Temperature Dependence of Damping and Frequency Shifts of the Scissors Mode of a trapped Bose-Einstein Condensate}
\author{Onofrio Marag\`o, Gerald Hechenblaikner, Eleanor Hodby, and Christopher Foot}
\address{Clarendon Laboratory, Department of Physics, University of Oxford,\\
Parks Road, Oxford, OX1 3PU, \\
United Kingdom.}
\date{\today}

\maketitle

\begin{abstract}
We have studied the properties of the scissors mode of a trapped
Bose-Einstein condensate of $^{87}$Rb atoms at finite temperature.
We measured a significant shift in the frequency of the mode below
the hydrodynamic limit and a strong dependence of the damping rate
as the temperature increased. We compared our damping rate results
to recent theoretical calculations for other observed collective
modes finding a fair agreement. From the frequency measurements we
deduce the moment of inertia of the gas and show that it is
quenched below the transition point, because of the superfluid
nature of the condensed gas.
\end{abstract}


The experimental discovery  of the scissors mode \cite{Bohle},
first predicted in a geometrical model \cite{nicola}, has been one
of the most exciting findings in nuclear physics during the last
two decades (see \cite{review} for a review). According to  the
geometrical  picture, such a mode arises from a counter-rotational
oscillation of the deformed proton and neutron fluids. Extensive
studies of this mode in the past two decades have investigated the
dependence of its strength on the nuclear deformation and the
relationship with the quadrupole collective mode and with the
superfluid moment of inertia of the nucleus \cite{review}.

Recently it has been possible to study the scissors mode in
Bose-Einstein condensates (BEC) of dilute gases \cite{Anderson}.
This transverse mode of excitation has been used to demonstrate
that the condensate can only flow in an irrotational manner
\cite{david,onofrio}. This work, together with the observation of
a critical velocity \cite{raman} and the occurrence of quantized
vortices \cite{matthews,madison} has provided conclusive evidence
for the superfluidity of a trapped condensate. The scissors mode
of a trapped BEC has been formally linked to the quadrupole
operator and to the moment of inertia of the superfluid system
\cite{francesca}.

In condensed gases the 'scissors' motion is an irrotational
oscillation of the condensate relative to the static trapping
potential. The deformation of the system can be controlled
independently using the trap frequencies (unlike nuclei). The most
striking difference between the scissors mode in nuclei and in
condensed gases is that the nuclear system is effectively at $T=0$
and the scissors excitation is fragmented because of the
interference of the orbital with the spin motion, whereas in a
trapped BEC it is possible to control the temperature of the gas,
and hence change the fraction $N_0/N$ of condensed atoms.

The spectroscopy of the low energy collective modes of the
condensate has been extended to include measurements of damping
rates and frequency shifts as a function of temperature
\cite{jin2,stamper}. These finite temperature measurements test
dynamical aspects of the BEC theory that require higher order
terms in the calculations. As yet the results have not been fully
explained \cite{hutchinson}.

In this paper we report a systematic study of the temperature
dependent damping and frequency shifts of the scissors mode
excitation in a trapped condensed gas. After a brief description
of the experimental procedure, we present the scissors mode data
for the frequency shifts and damping rates. We then compare our
results with the available theoretical calculations for other
collective modes. Finally we show how the scissors mode frequency
shifts are related to quenching of the moment of inertia of the
boson gas.

\begin{figure}
\centerline{\mbox{ \epsfxsize 3.8 in\epsfbox{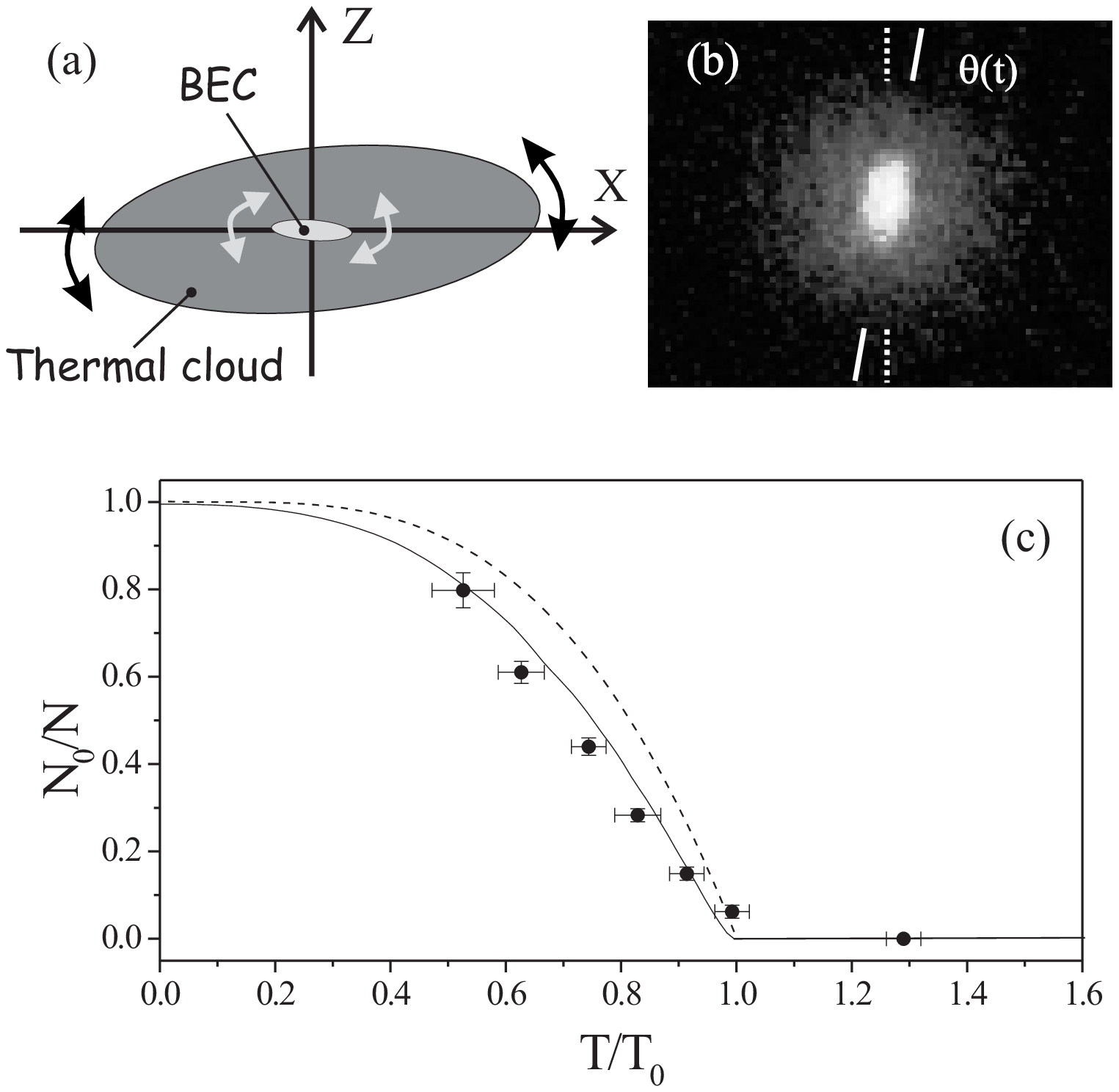}}}
\caption{(a) At finite temperature, during a scissors mode
oscillation, the condensate and the thermal cloud oscillate at
their characteristic frequencies. (b) Typical time of flight (TOF)
picture during a scissors mode oscillation at finite temperature:
the thermal cloud is isotropically expanded, while the condensate
has an elongated shape. From a 2D double Gaussian distribution fit
to the picture we extract the angle of the BEC. For this picture
$N_0/N=0.14$, $T/T_0=0.91$ and $\theta=-8.5^{\circ}$ with respect
to the vertical axis. (c) Condensate fraction as a function of the
reduced temperature $T/T_0$. The solid line is the interacting
theory prediction made by A. Minguzzi {\it et al.} for our
experimental conditions. The dashed line is the ideal trapped gas
cubic law prediction. The experimental points (solid circles) are
obtained from TOF images as explained in the text.} \label{finite}
\end{figure}

We excite the scissors mode using the technique described in our
previous paper \cite{onofrio}. In summary we prepare atoms at the
desired temperature $T$ in an untilted TOP trap
($\omega_x=\omega_y=126$ Hz, $\omega_z=\sqrt{8}\omega_x$). We then
adiabatically tilt the confining potential by a small angle
$\phi=3.4^{\circ}$ about the y axis (this reduces $\omega_z$ by
$\sim 2\%$). We then suddenly flip the trap angle to $-\phi$ to
excite the scissors mode in the xz plane with an amplitude
$\theta_0=2\phi$, about the new equilibrium position. The
condensate oscillates at a single frequency $\omega_{sc}$ (Fig.
\ref{finite} (a)). The low temperature limit of the condensate
frequency is the hydrodynamic value
$\omega_{sc}=\sqrt{\omega_z^2+\omega_x^2}$ \cite{david}. The
thermal component oscillates at two frequencies $\omega_+$ and
$\omega_-$, in the collisionless limit $\omega_+=\omega_z +
\omega_x$ and $\omega_-=\omega_z - \omega_x$. The high-lying
frequency $\omega_+$ corresponds to an irrotational quadrupole
velocity flow ($\nabla\times \vec{v} =0$) that is the classical
counterpart of the superfluid irrotational oscillation. The
low-lying frequency $\omega_-$ of the thermal cloud is related to
rotational flow, and the absence of a similar mode for the BEC is
the key feature that indicates superfluidity.

To observe the scissors mode oscillation for the thermal component
we imaged the atomic cloud in the trap using destructive
absorption imaging. The images of the cloud in the trap were
fitted with a 2D Gaussian distribution from which we extract the
cloud angle as a function of time (both above and below the
critical temperature).

To observe the condensate we switch off the trapping potential at
different times during the scissors mode oscillation, and allow
the cloud to expand for $14$ ms before taking an image. The
thermal cloud expands isotropically while the BEC component has an
elongated (prolate) shape (Fig. \ref{finite} (b)). The angle of
the expanded BEC component is extracted by fitting the two
component profile with a 2D double Gaussian distribution
\cite{note}. In this way we are able to separate the information
on the thermal cloud scissors mode and the BEC scissors mode.

The temperature of the ultracold atomic sample is extracted from a
2D Gaussian fit to the wings of the expanded thermal component
using the procedure well described in \cite{ensher}. For
temperatures below $0.5 T_0$ when the thermal cloud is not
visible, we deduced the temperature from the final rf by
extrapolating the best fit curve for the available temperature vs
rf cut frequency data. To determine the scaled temperature $T/T_0$
we need to calculate the critical temperature $T_0(N)$ at each
final rf cut - in evaporation each rf cut gives a different final
number of atoms $N$ and hence different $T_0(N)$. We calculated
$T_0(N)$ from the measured final number of atoms in each
time-of-flight picture and the known trap frequencies, using the
formula for a trapped boson gas \cite{temperature}. The
statistical uncertainty in $T/T_0$ is in the range $3-6\%$, but
increased to $10\%$ for $T/T_0 < 0.5$ where $T$ was extrapolated.
To check our temperature measurements we compared the condensate
fraction vs temperature with theoretical predictions for both the
ideal gas and interacting gas \cite{anna} (Fig. \ref{finite} (c)).

The measurements of the frequency and damping of the quadrupole
modes in previous work \cite{jin2,stamper} relied on measurement
of the radii of the condensate and the thermal cloud. These
measurements are sensitive to the fitting model and to shot to
shot noise because of their atom number dependence. However the
scissors mode has very well defined initial conditions for the
excitation of both the BEC and the thermal cloud and the angle
oscillations of the two components can be measured independently.
The angles of both condensed and thermal clouds do not depend on
atom number and could be fitted very reliably, even close to $T_0$
to give accurate measurements of frequencies and damping.

\begin{figure}
\centerline{\mbox{ \epsfxsize 3.2 in\epsfbox{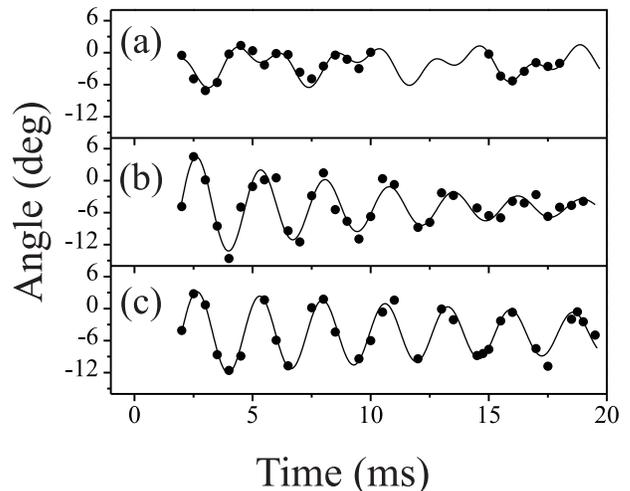}}}
\caption{Angle versus time during the scissors mode. (a) Thermal
cloud, $T/T_0=1.29$. The scissors mode is characterized by two
frequencies of oscillation with equal amplitude. (b) Condensate,
$T/T_0=0.74$. A heavily damped BEC oscillation is observed at a
single frequency, below that of hydrodynamic theory. (c)
Condensate, $T/T_0=0.53$. Lightly damped BEC oscillation at a
frequency that agrees well with the hydrodynamic value.}
\label{data}
\end{figure}

\begin{figure}
\centerline{\mbox{ \epsfxsize 4.0 in\epsfbox{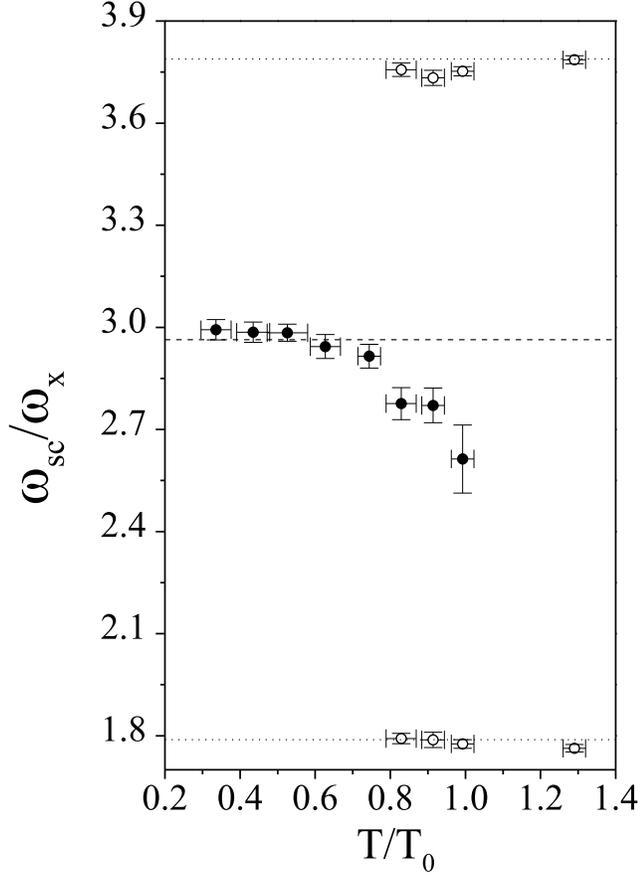}}}
\caption{Temperature dependent frequency shift of the scissors
excitation of the condensate (solid circles) and thermal component
(open circles). Significant negative shifts from the hydrodynamic
prediction (dashed line) of the condensate scissors mode frequency
are observed as the temperature increases. The low temperature
frequencies are systematically higher than the hydrodynamic value
by $\sim 1\%$ due to the finite number of atoms. The two
frequencies of the scissors mode of the thermal component do not
appear to be temperature dependent and are in good agreement with
collisionless predictions (dotted lines). The frequency spectrum
has been normalized by the radial trap frequency.} \label{freq}
\end{figure}

\begin{figure}
\centerline{\mbox{ \epsfxsize 3.2 in\epsfbox{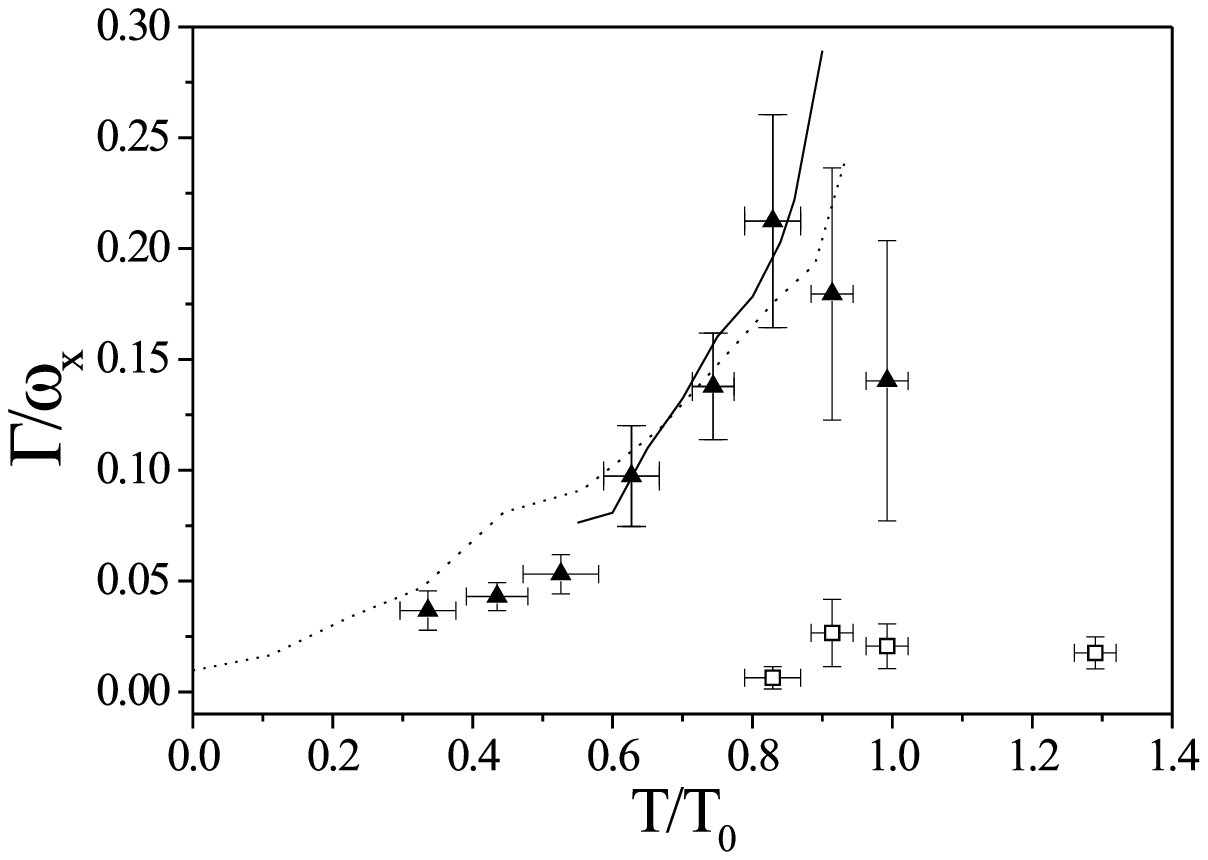}}}
\caption{The temperature dependence of the damping rate $\Gamma$
normalized to the radial trap frequency $\omega_x$ for the
condensate (solid triangles) and thermal component (open squares).
The solid line is the theoretical damping rate of the $|m=2|$ mode
calculated by Fedichev {\it et al.} (which agrees well with the
experimental data obtained at JILA for that mode) multiplied by
the frequency scaling factor $\omega_{sc}/\omega_{|m=2|}$ for the
scissors mode. The dotted line is the prediction for the $l=2$
mode by Rusch {\it et al.} in a spherical geometry rescaled by the
frequency factor for the scissors mode.} \label{damp}
\end{figure}

Figure \ref{data} shows three different scissors mode oscillations
at temperatures of (a) $T/T_0=1.29$, (b) $T/T_0=0.74$ and (c)
$T/T_0=0.53$. The measured frequencies of the thermal cloud do not
change with temperature and agree very well with the collisionless
frequencies, both above and below the critical temperature (see
also Fig. \ref{freq}). Both frequency components have the same
amplitude implying that energy is shared equally between
rotational and irrotational velocity flow. These thermal cloud
excitations have very small damping that is independent of
temperature, confirming the collisionless regime of the thermal
component in our experiment (see also Fig. \ref{damp}).

The scissors mode oscillation of the BEC component occurs at a
{\it single frequency} $\omega_{sc}$ (Fig. \ref{data} (c) and
(b)). However, the frequency of oscillation and its damping change
dramatically with temperature. The low temperature limit of the
frequency was $\sim 1\%$ higher than the hydrodynamic prediction
as shown in Fig. \ref{freq}. This agrees with the frequency we
calculated for finite number condensate ($2\times 10^4$) using the
method detailed in \cite{Pires}.

Figure \ref{damp} shows the experimental data for the temperature
dependent damping rates. Since a finite $T$ theory of the scissors
mode in a TOP trap is not yet available we have compared our
experimental results with recent theories for other observed
collective modes of a BEC.\\ Fedichev {\it et al.} \cite{damping}
have calculated the Landau damping rate for the $|m=2|$ and
low-lying $m=0$ modes of BEC in the usual TOP trap geometry
($\omega_x=\omega_y,\, \omega_z=\sqrt{8}\omega_x$) for the
experimental conditions of the JILA trap \cite{jin2}, that are
close to those of our experiment. In the region $0.5<T/T_0<0.9$
their theoretical calculations give a damping rate $\Gamma_{\nu}$
for the collective mode $\nu$ that is proportional to $T/\mu$ and
to the frequency of the excitation $\omega_{\nu}$:

\begin{equation}
\Gamma_{\nu}=A_{\nu} \omega_{\nu}\frac{T}{\mu} (n_0
a^3)^{1/2}\label{gamma}
\end{equation}

\noindent where $n_0$ is the peak density, $\mu$ is the chemical
potential and $A_{\nu}$ is a numerical coefficient dependent on
the shape of the collective excitation. For the quadrupole modes
they calculated $A_{|m=2|}\approx 7$ for the $|m=2|$ and
$A_{m=0}\approx 5$ for the low-lying m=0. The numerical
coefficient $A_{\nu}$ is likely to be similar for the scissors
mode. Assuming that $A_{sc}\approx A_{|m=2|}\approx 7$ and scaling
the results by Fedichev {\it et al.} \cite{damping} by the
frequency factor $\omega_{sc}/\omega_{|m=2|}$, we obtain the
theoretical damping rate for the scissors mode (Fig. \ref{damp},
solid line), which is in fair agreement with our data. This
numerical coefficient for the scissors mode could in principle be
evaluated by the method described by Fedichev {\it et al.}
\cite{damping}. They use an overlap of the wavefunction of the
mode with the wavefunctions of the thermal excitations following
classical trajectories within the range $\mu<T<0.9T_0$. This
method gives a way of calculating the damping process for a
trapped BEC, the evaluation of the trajectories and the stochastic
averaging is a numerical problem of moderate size. Our
measurements of the damping extend to temperatures where $T\leq
\mu$. In this region $0<T/T_0<0.5$ Fedichev {\it et al.} predict
damping proportional to $(T/\mu)^{3/2}$ \cite{damping} but our
experimental data have a weak dependence on the temperature.\\ In
a recent paper Rusch {\it et al.} \cite{martin} have used a full
second order theory to calculate the damping and frequency shift
of low energy collective modes for a trapped BEC in a spherical
geometry \cite{note2}. A spherical cloud has no scissors mode but
it is possible to identify this mode with the quadrupole mode
$l=2,m=\pm 1$. Thus we have used their calculations for the $l=2$
mode damping rate for comparison scaled by the frequency factor
$\omega_{sc}/\omega_{l=2}$ (Fig. \ref{damp}, dotted line). These
numerical calculations give a weakly linear dependence of damping
on temperature but the experimental data seem to follow a higher
power law.\\ We have not included the observed decrease in damping
at higher temperatures (above $0.85T_0$) in the above discussion
because the uncertainties are large for these data points. In
addition the frequency shift of the scissors mode at these
temperatures may lead to resonant coupling with other modes or
with the thermal cloud. For example in \cite{adams} the lowering
of the damping of the quadrupole modes close to the transition
point, has been interpreted as a dynamical coupling between the
condensed and thermal component.

The moment of inertia of a system is defined as linear response to
a rotational field. Superfluids have a moment of inertia $\Theta$
less than that of a classical rigid body $\Theta_{rig}$ with the
same density distribution. The scissors mode is ideally suited to
studying this quenching of the moment of inertia below $T_0$.
Zambelli and Stringari \cite{francesca} linked the reduced moment
of inertia of the boson gas
$\Theta^{\prime}=\Theta/\Theta_{\mathrm rig}$ to the quadrupole
moment Fourier signal $Q(\omega)$ for the condensate at $T=0$ and
the thermal cloud at $T>T_0$. Applying their derivation to the
scissors mode gives the moment of inertia of the condensate and
thermal clouds as:

\begin{equation}
\Theta_{\mathrm
cond}^{\prime}=\frac{(\omega_z^2-\omega_x^2)^2}{\omega_{sc}^4};\;
\Theta_{\mathrm
therm}^{\prime}=\frac{(\omega_z^2-\omega_x^2)^2}{\omega_+^2\
\omega_-^2}.\label{formula}
\end{equation}

\noindent We can use Eq. \ref{formula} to deduce the moment of
inertia for the condensate and thermal cloud separately. The
moment of inertia for the thermal component is consistent with the
rigid body value $\Theta_{\mathrm therm}^{\prime}=1$ for all
measured $T$. For the point at $0.99T_0$ in Fig. \ref{freq}, the
condensate scissors mode has a frequency such that its moment of
inertia is close to that of a rigid body. Whereas as $T
\rightarrow 0$ the condensate has a quenched irrotational value of
the moment of inertia characteristic of a superfluid. The lower
temperature points are in good agreement with the hydrodynamic
frequency and the reduced moment of inertia assumes the superfluid
value for our geometry $\Theta_{\mathrm
cond}^{\prime}=\epsilon^2=0.6$, where
$\epsilon=(\omega_z^2-\omega_x^2)/(\omega_z^2+\omega_x^2)$ is the
anisotropy of the trap \cite{francesca}.

In conclusion we have made precise measurements of the frequency
and damping of the scissors mode as a function of temperature. The
observed damping for low temperatures is consistent with that
expected from models of Landau damping, where the damping rate is
proportional to the mode frequency. At temperatures just below
$T_0$ the experimental damping rate was less than predicted. The
frequency shift as a function of temperature is consistent with a
quenching of the moment of inertia as the temperature is reduced
because of the superfluidity of the gas. We await a theoretical
calculation of this shift for comparison.

We would like to thank S. Stringari and F. Zambelli, N. Lo Iudice,
and all the members of the Oxford theoretical BEC group in
particular K. Burnett, M. Davis, M. Rusch, S. Morgan, for many
very useful discussions. We are also grateful to A. Minguzzi for
providing the theoretical prediction of Fig. \ref{finite} (c).

This work was supported by the EPSRC and the TMR `Cold Quantum
Gases' network program (No. HPRN-CT-2000-00125). O.M. Marag\`{o}
acknowledges the support of a Marie Curie Fellowship, TMR program
(No. ERB FMBI-CT98-3077).

\begin{references}
\bibitem{note} We have also used a 2D fitting distribution made of
a Gaussian profile for the thermal component and a Thomas-Fermi
profile for the condensate. This gave the same result for the
scissors angle, confirming that this is not affected by the choice
of the model to describe the cloud profile. The condensate
fraction was instead systematically lower of about $10\%$ compared
to the Gaussian profile fitting.
\end{references}
\end{document}